\documentclass{emulateapj}

\newcommand{\myemail}{sascha.quanz@astro.phys.ethz.ch}

\slugcomment{Published by ApJ Letters online February 28, 2013}

\shorttitle{A protoplanet candidate embedded in the circumstellar disk of HD100546}
\shortauthors{Quanz et al.}

\begin{document}


\title{A young protoplanet candidate embedded in the circumstellar disk of HD100546}

\author{Sascha P. Quanz$^{1,2}$, Adam Amara$^2$, Michael R. Meyer$^2$, Matthew A. Kenworthy$^3$, Markus Kasper$^4$, and Julien H. Girard$^5$}
\email{\myemail}

\altaffiltext{1}{Based on observations collected at the European Organisation for Astronomical Research in the Southern Hemisphere, Chile, under program number 087.C-0701(A).}
\altaffiltext{2}{Institute for Astronomy, ETH Zurich, Wolfgang-Pauli-Strasse 27, 8093 Zurich, Switzerland}
\altaffiltext{3}{Sterrewacht Leiden, P.O. Box 9513, Niels Bohrweg 2, 2300 RA Leiden, The Netherlands}  
\altaffiltext{4}{European Southern Observatory, Karl Schwarzschild Strasse, 2, 85748 Garching bei M\"unchen, Germany}      
\altaffiltext{5}{European Southern Observatory, Alonso de C\'ordova 3107, Vitacura, Cassilla 19001, Santiago, Chile}


\begin{abstract}
We present high-contrast observations of the circumstellar environment of the Herbig Ae/Be star HD100546. The final 3.8 $\mu m$ image reveals an emission source at a projected separation of 0.48$''\pm$0.04$''$ (corresponding to $\sim$47$\pm$4 AU) at a position angle of 8.9$^\circ \pm$0.9$^\circ$. The emission appears slightly extended with a point source component with an apparent magnitude of $13.2\pm0.4$ mag. The position of the source coincides with a local deficit in polarization fraction in near-infrared polarimetric imaging data, which probes the surface of the well-studied circumstellar disk of HD100546. This suggests a possible physical link between the emission source and the disk. Assuming a disk inclination of $\sim$47$^\circ$ the de-projected separation of the object is $\sim$68 AU. Assessing the likelihood of various scenarios we favor an interpretation of the available high-contrast data with a planet in the process of forming. Follow-up observations in the coming years can easily distinguish between the different possible scenarios empirically. If confirmed, HD100546 ``b" would be a unique laboratory to study the formation process of a new planetary system, with one giant planet currently forming in the disk and a second planet possibly orbiting in the disk gap at smaller separations.

\end{abstract}



\objectname{HD100546}


\section{Introduction}
To possibly extend the ongoing census of exoplanet demographics from our solar neighborhood to the entire Milky Way we need to understand planet formation and its dependence on initial physical and chemical conditions. A key step is to directly detect and characterize forming planets in their natal environment. 
Recently, based on sparse aperture masking observations, a few low-mass companion candidates have been revealed in the gap of their host star's transitional disks \citep[e.g., LkCa15 b, T Cha b;][]{kraus2012,huelamo2011}, but in some cases scattered light from the disk rim or other disk structures may be a valid explanation for the observed features \citep[e.g.,][]{cieza2013}. 
Until now no protoplanet has yet been found embedded in the optically thick gas and dust disk of its host star. Here we present high contrast imaging data revealing a protoplanet candidate embedded in the disk around the Herbig Ae/Be star HD100546.

\begin{deluxetable}{llc}
\centering
\tablecaption{Basic parameters of HD100546. 
\label{parameters}}           
\tablewidth{0pt}
\tablehead{
\colhead{Parameter} & \colhead{HD100546} & \colhead{Reference\tablenotemark{a}}
}
\startdata
RA (J2000) & 11$^h$33$^m$25$^s$.44  & (1) \\ 
DEC (J2000) & -70$^\circ$11$'$41$''$.24   & (1)\\
$J$ & 6.43$\pm 0.02$ mag  & (2)\\
$H$ & 5.96$\pm 0.03$ mag  &(2)\\
$K_s$ & 5.42$\pm 0.02$ mag & (2)\\
Mass & 2.4$\pm$0.1 M$_\sun$ & (3)\\
Age &  5...$>$10 Myr & (3),(4)\\
Distance & 97$^{+4}_{-4}$ pc & (1)\\
Sp. Type & B9Vne & (5)\\
\enddata
\tablenotetext{a}{(1) \citet{vanleeuwen2007}, (2) 2MASS point source catalog \citep{cutri2003}, (3) \citet{vandenancker1997}, (4) \citet{guimaraes2006}, (5) \citet{houk1975}.}
\end{deluxetable}

HD100546 (see Table~\ref{parameters} for stellar properties) has a complex circumstellar environment consisting of an inner disk from $\sim$0.2--4 AU, a disk gap from $\sim$4--13 AU, and a large outer disk from $\sim$13 AU out to a few hundred AU \citep[e.g.,][]{benisty2010,tatulli2011}. A massive planet was suggested to be orbiting in the gap \citep{bouwman2003}. Based on asymmetries in the line profile of [OI] \citep{acke2006} and OH \citep{liskowsky2012} dynamic evidence for such an object was found. In the case of OH the emission is thought to arise from an eccentric inner rim of the outer disk with the eccentricity being introduced by a planet. The outer disk has been resolved at multiple wavelengths including scattered light \citep[e.g.,][]{augereau2001,grady2001,ardila2007} where it shows peculiar features such as large-scale spiral arms. The remaining disk mass is estimated to be 10$^{-2}$ -- 10$^{-3}$ M$_\sun$ \citep[e.g.,][]{panic2010}. Recently, polarimetric differential imaging (PDI) in the near-infrared (NIR) revealed distinct sub-structures in the innermost few tens of AU of the disk \citep{quanz2011b}. 

\begin{deluxetable*}{lll}
\tablecaption{Summary of deep imaging observations in pupil tracking mode.
\label{observations}}           
\tablewidth{0pt}
\tablehead{
\colhead{Parameter}  & \colhead{HD100546} & \colhead{HD100546} \\
\colhead{} & \colhead{Hemisphere 1} & \colhead{Hemisphere 2} 
}
\startdata
Date & 2011-05-30 & 2011-07-13 \\
UT start & 22h:48m:18s & 23h:00m:49s  \\
UT end &  00h:14m:58s & 00h:26m:11s   \\
NDIT $\times$ DIT\tablenotemark{a}  & 200 $\times$ 0.15 s & 200 $\times$ 0.15 s \\
NINT\tablenotemark{b} & 130 & 126 \\
Parallactic angle start & -17.33$^{\circ}$ & 40.36$^{\circ}$ \\
Parallactic angle end & 10.33$^{\circ}$ & 63.93$^{\circ}$ \\
Airmass range& 1.45\dots1.43 & 1.54\dots1.75  \\
Mean DIMM seeing [$\lambda$=500 nm]&  0.6$''$ & 0.9$''$\\
$\langle \tau_0 \rangle_{\rm mean}$ / $\langle \tau_0 \rangle_{\rm min}$ / $\langle \tau_0 \rangle_{\rm max}$\tablenotemark{c} & 4.4 / 0.0 / 9.7 ms & 1.9 / 0.0 / 4.0 ms  \\
PA start\tablenotemark{d}  & -108.35$^{\circ}$ & 129.34$^{\circ}$ \\
PA end\tablenotemark{d}  & -80.86$^{\circ}$ & 153.04 $^{\circ}$
\enddata
\tablenotetext{a}{NDIT = Number of detector integration times (i.e., number of individual frames); DIT = Detector integration time (i.e., single frame exposure time).}
\tablenotetext{b}{NINT = Number of data cubes.}
\tablenotetext{c}{Average, minimum and maximum value of the coherence time of the atmosphere in data cube. Calculated by the Real Time Computer of the AO system.}
\tablenotetext{d}{Position angle of camera adaptor.}
\end{deluxetable*}

\section{Observations and data reduction}\label{observations_section}
HD100546 was observed with VLT/NACO \citep{lenzen2003,rousset2003} and its Apodozing Phase Plate (APP) coronagraph \citep{kenworthy2010}, which was already used in earlier exoplanet imaging projects \citep{quanz2010,quanz2011a,kenworthy2013}. 
We used the L27 camera ($\sim$ 27 mas pixel$^{-1}$) with the L$'$ filter ($\lambda_{c}=3.8\,\mu$m, $\Delta\lambda=0.62\,\mu$m) in angular differential imaging mode \citep[ADI; ][]{marois2006}. Two datasets were taken on two different nights, referred to as``hemisphere 1" and ``hemisphere 2", with a 180$^\circ$ offset in position angle of the APP to allow its high-contrast half to cover most of the circumstellar environment. Using the ``cube mode" option, all exposures, each 0.15 s long, were saved individually. The core of the stellar PSF was slightly saturated. Data cubes consisting of 200 exposures were taken using a 3--point dither pattern along the detectors x-axis with roughly 7$''$ separation. Table~\ref{observations} summarizes the observations and the observing conditions.

For photometric calibration we observed HD100546 in the NB3.74 filter ($\lambda_{c}= 3.74\,\mu$m, $\Delta\lambda=0.02\,\mu$m; centered on the Pf$_\gamma$ line). The observing strategy was identical to the deep science observations, but these exposures were in the linear detector regime (0.2 s exposure time) and only 2 data cubes with 150 exposures each were recorded. No photometric or astrometric standard star was observed. 

The basic data reduction steps (bad pixel cleaning, sky subtraction, image alignment) were done in a similar way as described in \citet{quanz2010}. During the alignment process the images were re-binned to twice their original resolution. From each image in the stack of aligned exposures we created a 2$''\times$2$''$ sub-image centered on the star. Individual images showing bad AO correction or not covering the full size of the sub-images 
were disregarded. In the end we had a stack of 16,117 images for hemisphere 1 and 18,916 images for hemisphere 2.

The PSF-subtraction was done using the principle component analysis based software package {\sc PynPoint} \citep{amaraquanz2012}, and the results were confirmed by using the LOCI algorithm \citep{lafreniere2007b}. For the final {\sc PynPoint} images we used 80 PCA coefficients and kept only the best 12000 images in terms of total residuals over the whole image frame. After PSF subtraction, each image was de-rotated to the same field-orientation and we computed the mean image of the image stack clipping data points that were beyond 2.5$\sigma$ of the mean. The results shown below are robust against all of these numbers (see next section).

\begin{figure*}
\centering
\plotone{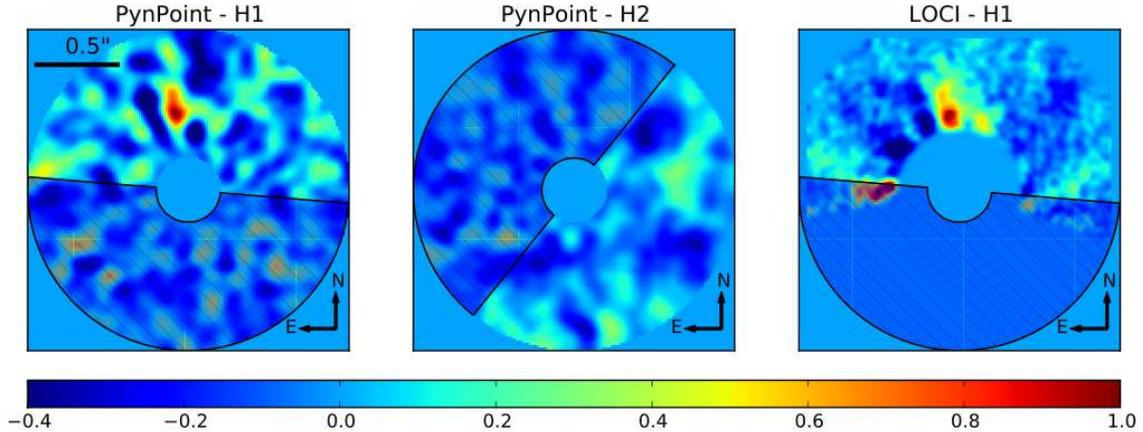}
\caption{NACO/APP L$'$ images of the circumstellar environment of HD100546. From left to right: Final {\sc PynPoint} images of hemisphere 1 and hemisphere 2 and final LOCI image of hemisphere 1. An emission source is clearly detected in left and right panel. The shaded area indicates the region that was only covered by the low sensitivity hemisphere of the APP. The images are scaled with respect to their peak flux.
\label{images}}
\end{figure*}

For LOCI we median combined 20 consecutive exposures into a single image in those data cubes where this was possible, resulting in 734 stacked images. The LOCI algorithm was then applied to this stack of images using the following LOCI parameters: FWHM=8 px, $N_\delta$=0.75, $dr$=8, $N_A$=500. The choice of these values reflect the fact that the images have been re-binned to twice their original resolution (see above). All final images (see, Figure~\ref{images}) were smoothed with a circular gaussian with a width of 3 pixels. 

\section{Results and analysis}\label{results}
\subsection{Detection of an emission source}
In Figure~\ref{images} an emission source is revealed north of the central star in the hemisphere 1 dataset. To examine the robustness of this detection we did a series of tests:
\begin{itemize}
\item We varied the number of PCA coefficients used in {\sc PynPoint} (between 20 and 200). 
\item We split the dataset in half in different ways including random selections of $\sim$50\% of the images.
\item We applied the LOCI algorithm as independent reduction approach.
\end{itemize}
In all cases we found a bright feature at the same location. In addition, we analyzed the hemisphere 2 dataset in exactly the same way without finding a persistent source at any location in the final image.

Using the approach described in \citet{quanz2011a} and a 8-pixels wide aperture ($\sim 1\cdot$FWHM), the source has a signal-to-noise of $\sim$15 in the final LOCI image. The detected emission appears slightly elongated in the northern direction (Figure~\ref{images}). To estimate the photometry and astrometry of the source we did a detailed analysis inserting fake negative planets (details see below). It showed that the observed emission can be explained with a point source plus some extended component. For the point source the projected separation amounts to $\sim$0.48$''\pm0.04''$ ($\sim 47\pm4$ AU). The uncertainties in the exact location of the central star and the point source component in x and y on the detector were 0.5 and 1 pixel, respectively. The position angle of the point source with respect to the star is $\sim$8.9$^\circ\pm$0.9$^\circ$. This error excludes any systemic error in the orientation of the camera with respect to the true celestial north, which is estimated to be $\lesssim$0.5$^\circ$ based on calibration data from an ongoing large imaging program (PI: J.-L. Beuzit). 

To estimate the brightness of the point source we inserted artificial negative planets in the individual exposures and re-ran {\sc PynPoint}. 
For the fake objects we used an unsaturated PSF of HD100546 from one of the photometric calibration datasets. To scale the flux of these objects, the difference in exposure time between the science and the calibration images had to be considered as well as the transmission curves of the two different filters\footnote{see, http://www.eso.org/sci/facilities/paranal/instruments/\quad naco/inst/filters.html for transmission curves of the NACO filters.}. Using published L-band spectra for HD100546 from ISO and VLT/ISAAC \citep[][and references therein]{geers2007} we derived a throughput fraction of $\sim$$0.074\pm0.002$ for the narrow band filter compared to the broadband L$'$ filter. The error arises from changes in the Pf$_\gamma$ line emission in HD100546 between the two datasets suggesting that the NB3.74 filter traces variable accretion activity. Also the whole NIR and MIR continuum varies with an offset of a factor of $\sim$1.25 between the two datasets of \citet{geers2007}, which impacts the error in the final photometry. 

For our contrast estimates we varied the brightness of the injected fake sources in steps of 0.1 mag and used two independent methods. First, we searched for a negative source that, when subtracted, yielded a remaining flux at the object's location similar to the flux level in the surroundings, i.e., in the extended flux component. Secondly, we canceled out all the flux at the object's location. Our best estimate contrasts were  $\Delta$L$' = 9.0 \pm 0.3$ mag and  $\Delta$L$' = 8.3 \pm 0.3$ mag for both methods, respectively. A contrast closer  $\Delta$L$'\approx 9.0$ mag appears to be more likely, as strong residuals become present in the vicinity of the object -- not at the object's location itself -- if we cancel out its peak flux completely. For the rest of the analyses and discussion we use $\Delta$L$'\approx9.0$ mag as default value. 
The key points of the discussion and conclusions remain unchanged for a smaller value of the contrast.

\begin{figure*}
\centering
\epsscale{1.1}
\plotone{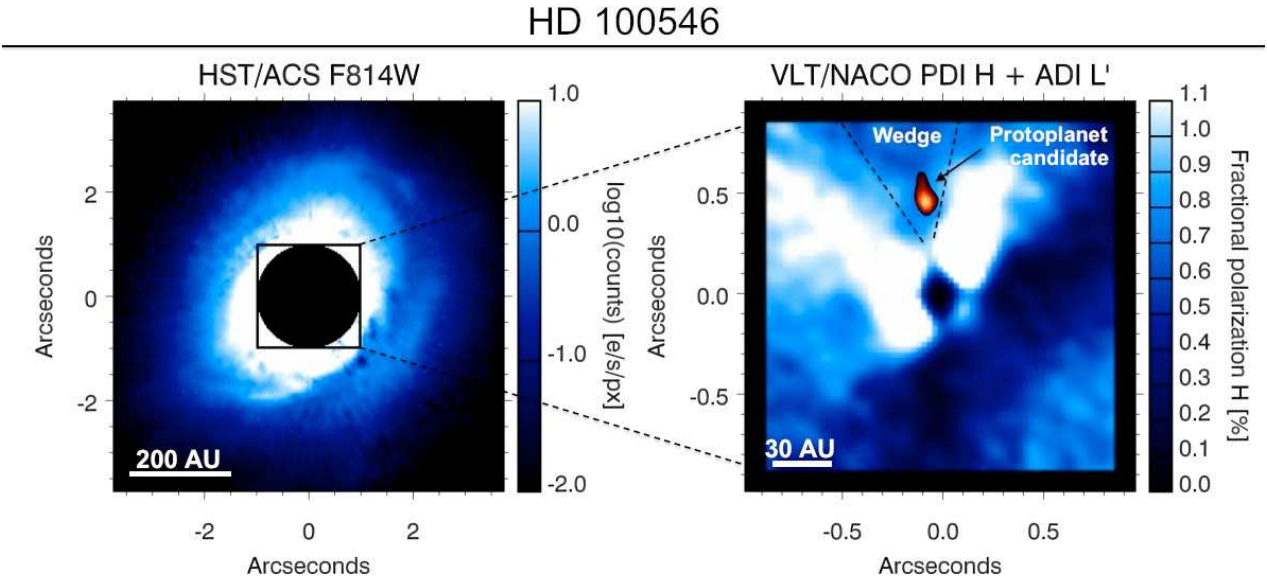}
\caption{The HD100546 disk on different scales. In the HST/ACS image obtained in the F814W filter (left) the circumsteller disk around HD100546 can be traced out to a few hundred AU in scattered light \citep{ardila2007}. The inner disk regions ($\sim$1$''$ in radius) are hidden behind the coronagraph or suffer from PSF subtraction residuals. The polarization fraction image (left) obtained at the VLT in PDI mode in the H band \citep{quanz2011b} probes regions very close to the star, enabling the detection of disk asymmetries not accessible with other imaging techniques. The position of the planet candidate is overlaid in the PDI image. North is up and east to the left in both images. \label{disk}}
\end{figure*}

The observed flux densities for HD100546 based on the spectra mentioned above translate into an apparent brightness of L$'\approx 4.1 \dots$ 4.3 mag, which fits well to the L$=4.02 \pm 0.06$ mag reported in \citet{dewinter2001}. Hence we derive an apparent magnitude of L$'$=13.2 $\pm$ 0.4 mag for the point source component. Here the error is the root of sum of squares of 0.3 mag for the uncertainty in the contrast and $\sim$0.25 mag for the uncertainties in the MIR continuum. Compared to these uncertainties the variability of the Pf$_\gamma$ line flux and the intrinsic error in our photometric observations are negligible.

\subsection{Estimating the minimum luminosity} 
Assuming that the flux of the point source peaks in the L$'$ filter we can estimate its blackbody temperature using Wien's law. We can then derive a lower limit on the object's luminosity by taking into account its apparent L$'$ magnitude and its distance. 
Integrating over all frequencies this exercise yields a minimum luminosity of $L \gtrsim 4\cdot 10^{-4} L_\sun$. 

\subsection{Interaction with  the circumstellar disk?}

The VLT/NACO PDI data presented in \citet{quanz2011b} have sufficient spatial resolution and inner working angle to probe the disk surface on scales relevant for the APP dataset. Those NIR observations revealed sub-structures in the disk in the inner few tens of AU. In particular the existence of a disk ``hole" was suggested as both the final polarization intensity images as well as the polarization fraction images in H and K$_s$ revealed a local flux deficit at the same location.

In Figure~\ref{disk} we show the large scale disk environment revealed by HST/ACS \citep{ardila2007} and then, zooming in the inner disk regions, the polarization fraction image of the PDI study. We  overplot in red the contours of the object detected here. The disk is inclined by $\sim$47$^\circ\pm$3$^\circ$ and the position angle of the disk major axis is $\sim$138$^\circ\pm$4$^\circ$ \citep{quanz2011b}.
If the disk surface was smooth and azimuthally symmetric, the disk image shown in Figure~\ref{disk} should be mirror symmetric with respect to the disk minor axis running with a position angle of $\sim$48$^\circ$ through the image center \citep{quanz2011b}. However, there are clear asymmetries in form of a deficit in polarization fraction in northern direction, i.e., along the position of the detected object. Based on Figure~\ref{disk} the disk ``hole" extends to larger separations and appears more like a ``wedge". As discussed in \citet{quanz2011b} the underlying physical reason for this feature is not clear at the moment (e.g., drop in surface density, disk surface geometry, changing dust properties). However, finding an asymmetry at this specific location renders plausible a physical link between those structures and the source detected here.

\section{Discussion}
\subsection{The image of an embedded exoplanet?}

Based on the object's position angle, the disk inclination and the distance to HD100546, the object's de-projected separation from the central star is $\sim$68$\pm$10 AU, i.e., within the large circumstellar disk. Different scenarios to explain both the L band emission and the observed disk structure can be assessed:

\emph{Background source:} A background source would be observed through the HD100546 disk. Based on the disk model presented in \citet{mulders2011} background flux in the L band should be attenuated by a factor of $\sim$6.7$\cdot10^{-3}\approx5.4$ mag at a location of $\sim$70 AU\footnote{This factor does not include that the object is seen through an inclined disk which would yield an even higher optical depth.}. Taking this factor into account we used the Besancon galactic model \citep{robin2003}  to estimate the number of objects in the apparent magnitude range 7 mag $\leq$ L $\leq$ 9 mag. This yielded $\sim$330 objects in a 2 square degree patch on the sky centered around HD100546. This number translates into a probability of having such a physically unrelated source in a 1$''$ $\times$ 1$''$ field of view around the star of $p = 1.3 \cdot 10^{-5}$. Furthermore, the fact that the L band emission appears to be extended argues against a background object.

\emph{Disk feature:} The observed L$'$ brightness and minimum luminosity are difficult to explain with disk-internal processes alone as the expected temperature in the disk mid-plane at the location of the source is only $\sim$50 K \citep{mulders2011}. Furthermore, we are not aware of shock-processes that act only locally and might lead to the observed luminosity in a disk that appears to be not very massive. If it was scattered light that we see, one would expect that also in the NIR a maximum in scattered light would be seen. Using the PDI images as tracer for scattered light we find a local \emph{minimum} here as described above. 

\emph{Photospheric emission:} If the observed point source flux arose solely from the photosphere of a young object the COND and DUSTY models suggest masses between $\sim$15 -- 20 M$_{\rm Jupiter}$ for an age of 5 -- 10 Myr \citep{baraffe2003, chabrier2000}. Models with lower specific entropy in the initial conditions for the formation process predict even higher masses \citep[cf.][]{spiegel2012}. Classical binary formation via core fragmentation or formation via disk instability when the disk was still massive would be the preferred formation mechanisms for an object of this mass. In this case the object formed roughly coeval with the star and would have had time to significantly alter the structure of the main disk, e.g., dynamically clearing a large azimuthal gap, which has not been observed. 

\emph{Ejected planet:} Another massive planet is thought to be orbiting in the inner disk gap \citep[e.g.,][]{acke2006,tatulli2011} and we speculate that dynamical interactions between multiple planets and the disk could have led to an ejection event. 
The emission we see in the L$'$ images would then be a combination of the planet's intrinsic emission plus extra luminosity from disk material being heated from the planet moving through the disk. Assuming that the planet was initially orbiting at 10 AU its orbital period was $\sim$20 years yielding an orbital velocity of $\sim$3.2 AU$\,$yr$^{-1}$ ($\sim$15.2 km$\,$s$^{-1}$). If the ejection velocity is a few times that value it would have taken the planet less then 20 years to reach its current location and within less than 100 years it would be beyond the extent of the observable disk. Given the age of the system, this timescale is extremely small and observing the object exactly at the right time is unlikely. Adding further complexity to this scenario, the ejection needs to occur roughly in the plane of the disk to make a link to the observed disk structures. 
 
\emph{Forming planet:} In our view the best explanation for the observed morphology of both the disk and the emission source is the detection of a planet during its formation process. The luminosity of the object is not coming from an isolated photosphere, but rather the planet is still accreting material from the disk. Young, forming gas giants with masses between one and a few Jupiter masses are expected to have luminosities between 10$^{-4}$--10$^{-2}$$L_\sun$ during the first few hundred thousand years after gas runaway accretion sets in \citep{mordasini2012}, in agreement with our lower limit. Furthermore, an object in this mass range is expected to affect the disk structure much less and an azimuthal gap -- if it exists -- might be below our detection limits in the PDI data. A narrow gap is hard to see in an inclined disk. 
This scenario could also explain the extended component of the emission detected here with some disk material being heated in the accretion process similar to the case of LkCa15 b 

\citep[see][for possible mechanisms to heat the surrounding material during the accretion process]{kraus2012}. However, we acknowledge that from a theoretical perspective the formation of a gas giant planet at this location is not readily explainable using \emph{first principles}. For core accretion the timescales to assemble a massive rocky core seem to exceed the estimated age of the star, and given the observed disk parameters the disk does not seem to be gravitationally unstable.

\subsection{Observational tests to distinguish the scenarios}
The different dynamics involved in the scenarios outlined above and multi-wavelength photometry and/or spectra will eventually help us to confirm which hypothesis is correct. The proper and parallactic motion of HD100546 will allow us to rule out a background source with second epoch observations obtained as early as mid 2013. To distinguish between the ejection scenario and the ``forming planet" scenario, the baseline for follow-up observations needs to be a few years. While the object is expected to orbit its star with a period of $\sim$360 years in the latter case, it should move away quickly in radial direction if it were ejected.  
Also, high spatial resolution ALMA observations will help to search for an azimuthal gap in the surface mass density (gas and/or dust) at the planet's location. Spatially resolved information about the existence (or non-existence) of an azimuthal disk gap may allow us to derive a dynamical mass estimate for the planet \citep{lin1986,bryden1999}.

\section{Conclusion}
We have presented observational evidence that a gas giant planet could be forming in the circumstellar disk around the Herbig Ae/Be star HD100546 at a separation of $\sim$68 AU. This scenario, among others that we discussed, seems best capable of explaining most of the available data. However, some aspects remain qualitative and follow-up observations are required to validate our proposed interpretation. Together with LkCa15 b \citep{kraus2012}, the object presented here is currently the best candidate for a forming young gas giant planet. Particularly interesting is that  HD100546 ``b" would be the first protoplanet that is still embedded in a circumstellar disk and that it forms at large orbital separations. If confirmed, HD100546 would be a unique laboratory to study planet formation and the interaction between forming planets and the disk directly.

\acknowledgments
This research made use of the SIMBAD database, operated at CDS, Strasbourg, France, and of NASA's Astrophysics Data System. V. Geers kindly provided us with the ISO and ISAAC spectra. We are indebted to F. Meru, C. Dominik, H. M. Schmid and R. v. Boekel for helpful discussions. 



{\it Facilities:} \facility{VLT:Yepun (NACO)}

\end{document}